\documentclass{article}

\usepackage{arxiv}

\usepackage{microtype}
\usepackage{graphicx}
\usepackage{subfigure}
\usepackage{booktabs} 
\usepackage{comment}

\usepackage{hyperref}

\usepackage{amsmath}
\usepackage{amssymb}
\usepackage{mathtools}
\usepackage{amsthm}

\usepackage{algorithm}
\usepackage{algorithmic}

\usepackage[capitalize,noabbrev]{cleveref}

\theoremstyle{plain}

\newtheorem*{example}{Example}

\usepackage[textsize=tiny]{todonotes}

\renewcommand{\eqref}[1]{(\ref{#1})}

\usepackage{enumitem}
\usepackage{bm}
\usepackage{bbm}
\usepackage{multirow}
\usepackage{natbib}
\allowdisplaybreaks

\usepackage{xcolor}
\newcount\Comments  
\newcommand{\kibitz}[2]{\ifnum\Comments=1\textcolor{#1}{#2}\fi}

\usepackage{url}

\title{General Bayesian Predictive Synthesis}

\author{%
  Masahiro Kato \\
  Data Analytics Department\\
  Mizuho-DL Financial Technology Co., Ltd.\\
  Tokyo, Japan \\
  \texttt{masahiro-kato@fintec.co.jp} \\
}

\begin{document}

\maketitle

\begin{abstract}
This study investigates Bayesian ensemble learning for improving the quality of decision-making. We consider a decision-maker who selects an action from a set of candidates based on a policy trained using observations. In our setting, we assume the existence of experts who provide predictive distributions based on their own policies. Our goal is to integrate these predictive distributions within the Bayesian framework. Our proposed method, which we refer to as General Bayesian Predictive Synthesis (GBPS), is characterized by a loss minimization framework and does not rely on parameter estimation, unlike existing studies. Inspired by Bayesian predictive synthesis and general Bayes frameworks, we evaluate the performance of our proposed method through simulation studies.
\end{abstract}

\section{Introduction}
This study aims to develop a Bayesian decision-making method with experts' advice. We consider a situation where a decision-maker chooses an action and incurs a loss corresponding to the chosen action. To make better decisions, the decision-maker utilizes experts' advice, provided as the experts' predictive distribution of losses they incur when following their action choices. Our proposed method relies on a Bayesian framework that incorporates this distributional information into the decision-making process. We focus on making better decisions with lower loss rather than on point estimation of parameters.

Although our framework is novel, it relates to various closely related fields. One such field is ensemble learning, which has garnered attention for its empirical success \citep{Tibshirani1996,Lakshminarayanan2017,Osband2016}. Another related field is the problem of prediction with experts' advice for decision-making, extensively studied in both offline and online learning \citep{Cesa-Bianchi_Lugosi_2006,Vovk1998,Freund1997}.

\begin{figure}[t]
    \centering
    \includegraphics[width=130mm]{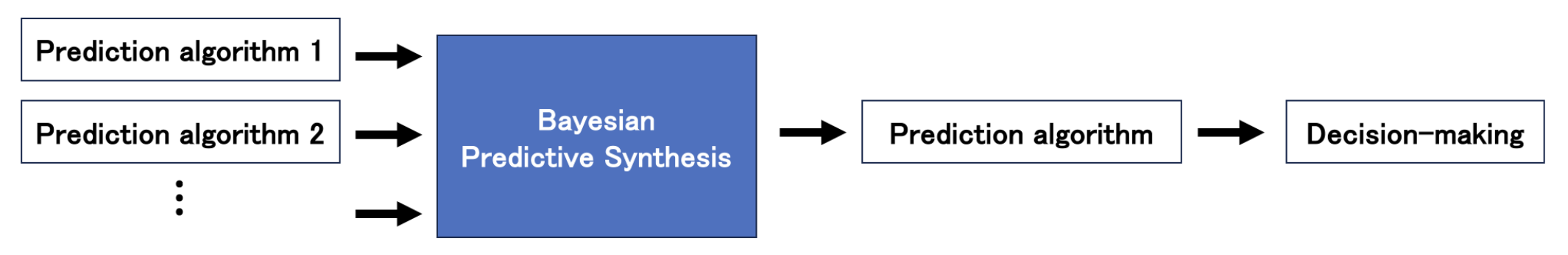}
    \vspace{-0.3cm}
    \caption{The concept of BPS: decision-making via parameter estimation.}
\label{fig:point}
    \vspace{0.3cm}
    \includegraphics[width=130mm]{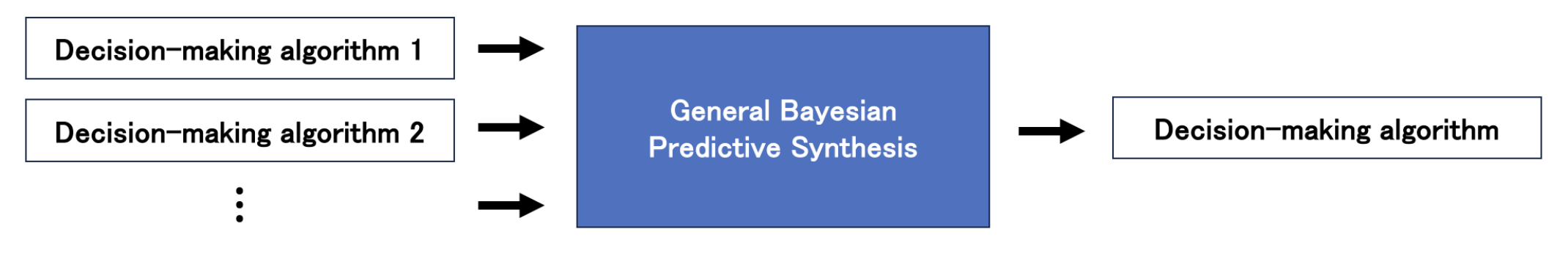}
    \caption{The concept of GBPS: direct decision-making.}
    \vspace{-0.3cm}
\label{fig:decision}
\end{figure}

Existing approaches are mostly based on frequentist or adversarial settings. In contrast, Bayesian modeling is crucial for incorporating uncertainty and agents' opinions to make better decisions. However, despite extensive studies in related areas, a Bayesian framework for this problem has not been sufficiently explored \citep{Gal2016,Pearce2020}. Pioneering approaches include Bayesian \emph{agent/expert opinion analysis}, also known as Bayesian predictive synthesis (BPS) \citep{Lindley1979,West1984,GenestSchervish1985,West1988,West1992}. The BPS has been developed into a more general framework, considering various applications, including time series analysis \citep{mcalinn2019dynamic}. Although these studies significantly contribute to the field, their focus lies in parameter estimation and does not directly address decision-making (Figure~\ref{fig:point}).

For example, \citet{tallman2023bayesian} considers decision-making after Bayesian prediction of parameters of interest, effectively bypassing an intermediate problem to solve the problem of interest. However, this approach can be further improved by directly optimizing the decision-making problem. This learning strategy is known as Vapnik's principle \citep{Vapnik2006}, which states, ``When solving a problem of interest, do not solve a more general problem as an intermediate step.''

In this study, we propose a Bayesian framework for decision-making. In Section~\ref{sec:problem}, we define the prediction problem with experts' advice and propose General BPS (GBPS) methods for both static and dynamic settings. In a static setting, we have independent and identically distributed (i.i.d.) observations and train a policy. In a dynamic setting, also referred to as an online setting, we observe data sequentially and make decisions across multiple periods. We introduce policy learning in causal inference for the static setting and portfolio optimization for the dynamic setting as examples. Our method incorporates experts' advice into the decision-making process (Figure~\ref{fig:decision}). In the empirical evaluation, we investigate the performance in portfolio optimization (Section~\ref{sec:application}).

Our framework and method are inspired by existing algorithms for the experts' problem and the general Bayes framework. In general Bayes, we directly address the decision-making problem with the idea of loss minimization instead of parameter estimation \citep{Bissiri2016}. Since our method is inspired by general Bayes, we refer to it as GBPS.

In a dynamic setting, we employ dynamic linear models with the state-space approach. This choice follows \citet{mcalinn2019dynamic}, which proposes a BPS method that utilizes dynamic linear models \citep{WestHarrison1997book2,Prado2010}. Independently of us, a state-space approach has also recently been introduced by \citet{luxenberg2024exponentially}, considered a frequentist or adversarial counterpart to our study.

\section{Basic Formulation of GBPS}
\label{sec:problem}
In this section, we describe our basic formulation.

\subsection{Action and Loss}
There is a decision-maker who chooses an action from a set of $K$ actions. After choosing an action, the decision-maker immediately incurs a loss corresponding to the chosen action.

If the loss only depends on the actions, we define it as
\[\ell: [K] \to \mathbb{R},\]
where $[K] \coloneqq \{1,2,\dots,K\}$. 

For an integer $n \in \mathbb{N}$, let $\Delta^n \coloneqq \left\{\left(u(a)\right)_{i\in[n]}\in[0, 1]^K \colon \sum_{i\in[n]}u(i) = 1\right\}$ be a probability simplex. The decision-maker chooses an action $a$ with probability $\pi(a) \in \Delta^K$, referred to as a policy.

Our goal is to train a policy $\pi \in \Delta^K$ to minimize the expected loss:
\begin{align}
    \mathbb{E}\left[\sum_{a\in[K]}\pi(a)\ell(a)\right],
\end{align}
where $\mathbb{E}[\cdot]$ denotes the expectation operator over all related randomness, including the policy.

If we consider the reward-maximization problem, we can multiply the loss by minus one and interpret our problem as a reward-maximization problem. 

This study restricts the class of policies that we can train. We consider a policy represented by a weighted sum of the experts' suggested policies. We delve into this formulation in the subsequent section. This setting implies that we train a policy as an ensemble of the experts' policies. 

\begin{example}[Policy learning]
Consider policy learning in causal inference. Assume that there are $K$ treatments. A decision-maker can observe $n$ observations defined as $\big\{(X_i, A_i, Y_i)\big\}^n_{i=1}$, where $X_i \in \mathbb{R}^d$ is a $d$-dimensional covariate, $A_i \in [K]$ is an indicator of treatment, and $Y_i$ is an outcome corresponding to $A_i$. By using potential outcomes $\{Y_i(a)\}_{a\in[K]}$, we denote the observed outcome as $Y_i = \sum_{a\in[K]}\mathbbm{1}[A_i = a] Y_i(a)$. The decision-maker's goal is to obtain a policy $\pi$ that maximizes 
\[\mathbb{E}_X\left[\sum_{a\in[K]}\pi(a\mid X_i)\mu(a\mid X)\right],\]
where $\mathbb{E}_X$ is an expectation over $X$ and $\mu(a\mid X) \coloneqq \mathbb{E}\big[Y(a)\mid X\big]$ is the expected outcome of $Y_i$ conditional on $X$. Note that we cannot observe $\{Y_i(a)\}_{a\in[K]}$ directly since we can only observe $Y_i(A_i)$. Therefore, we need to estimate $\mathbb{E}[Y_i(a)\mid X_i]$ in some way. There are typically three approaches. The first approach is to estimate $\mu(a\mid X_i)$ using regression, known as the direct method (DM). The second approach is to employ the inverse probability weight (IPW) estimator as $\sum_{a\in[K]}\frac{\mathbbm{1}[A_i = a]Y_i}{\pi(a\mid X_i)}$, where $\pi(a\mid X_i) = \mathbb{P}\left(A_i = a\mid X_i\right)$ is the propensity score. In our framework for policy learning, we set $\ell(a, z_i) = - \widehat{\mu}(a\mid X_i)$, where $\widehat{\mu}(a\mid X_i)$ is an estimator of $\mu(a\mid X_i)$.
\end{example}

\subsection{Experts}
To train a policy, we assume the availability of $J$ experts who provide their own policies to assist our decision-making. This setting is common in online learning as a problem with experts' advice. Unlike existing studies, we further assume the availability of the predictive distributions of $J$ experts' policies; that is, the policies are realizations generated from the decision-makers' predictive distributions. 

We define a policy of each expert $j\in[J]$. Let $w_{j}(z) \coloneqq \left(w_j(a)\right)_{a\in[K]} \in \Delta^K$ be expert $j$'s policy. If the decision-maker follows this policy, they incur a loss 
\begin{align}
    z_j \coloneqq \sum_{a\in[K]}w_j(a)\ell\left(a\right).
\end{align}
We also let $z = (z_j)_{j\in[J]}$ be the set of the experts' policies. 

The experts provide their losses as predictive distributions, not just single $w_j$. We define expert $j$'s predictive distribution of $w_j$ as $h_{j}(z_j)$.

\subsection{Ensemble Learning}
In this study, we train a new policy by ensembling the experts' policies. We define our policy as a weighted sum of the experts' policies and then aim to train the weights.

Formally, let $\theta \in \Delta^J$ be an ensemble parameter, and let $\pi^{\theta, w}$ be a new policy defined as
\begin{align}
    \pi^{\theta, w}(a) = \sum_{j\in[J]}\theta_jw_j(a).
\end{align}

Under this new policy $\pi^{\theta, w}(a)$, we incur the following loss:
\begin{align}
    \sum_{j\in[J]}\theta_j\sum_{a\in[K]}w_j(a)\ell(a) = \sum_{j\in[J]}\theta_jz_j,
\end{align}
which is denoted by $L(\theta, z)$.

We train $\theta$ to minimize the loss $\mathbb{E}[L(\theta, z)]$. To train $\theta$, we employ the Bayesian predictive synthesis framework and generalize it as a method for decision-making.

\subsection{GBPS}
The GBPS is a framework to train a new policy by ensembling experts' policies, employing a Bayesian approach with the experts' predictive distributions.

In the GBPS, we obtain the decision-making parameter $\theta$ via the following posterior distribution:
\begin{align}
    \pi(\theta \mid \mathcal{H}) \propto \int \cdots \int \exp\left(-L(\theta, z)\right)\prod_{j\in[J]}h_j(z_j) \, \mathrm{d}z_j,
\end{align}
where $\mathcal{H} = (h_j)_{j\in[J]}$ denotes the set of the predictive distributions.

This posterior distribution is an extension of the BPS. We justify its use in Section~\ref{sec:justify}.

Our method also shares motivation with general Bayes \citep{Bissiri2016} and Bayesian model averaging (BMA).

The above formulation is a basic one, and we can develop variants from it. In Section~\ref{sec:time_series}, we introduce a combination with the state-space model using dynamic linear models.

\subsection{Justification}
\label{sec:justify}
Bayesian ensemble learning has been developed by extending the BMA \citep{Monteith2011}. 

\paragraph{BPS.} We consider a problem where we are interested in predicting a variable $y\in\mathbb{R}$ by using the expert $j$'s prediction $w_j$ with its predictive distribution $h_j(z_j).$ 

Under the BPS framework, the predictive distribution of $y$ is given as
\begin{align}
    &\pi\big(y \mid \theta\,; \mathcal{H}\big) \coloneqq \int \alpha\big(y\mid \theta, z\big) 
\prod_{j\in[J]}h_{j}(z_j)\mathrm{d}z_j. 
\end{align}

Then, the posterior distribution of $\theta$ is given as
\begin{align}
    &\pi\big(\theta \mid y\,; \mathcal{H}\big) = \frac{\pi\big(y\mid \theta\,; \mathcal{H} \big)\pi\big(\theta\,;  \mathcal{H} \big)}{\pi\big(y\,; \mathcal{H}\big)},
\end{align}
where $\alpha$ is a synthesis function and $\theta\in\Theta$ is a parameter of the synthesis function. The synthesis function synthesizes $z = (z_j)_{j\in[J]}$ generated from $\mathcal{H} = (h_j(z_j))_{j\in[J]}.$ 

\paragraph{Linear models with a Gaussian error.}
A typical choice of the synthesis function is to assume a linear model. We consider predicting $y$ by a linear regression model with a Gaussian error defined as
\begin{align}
    y = \langle \theta, z \rangle + \epsilon\quad \epsilon \sim \mathcal{N}(0, \sigma^2),
\end{align}
where $\sigma^2$ is a variance. Although we assumed that the variance $\sigma^2$ is known, it can be estimated in BPS, as well as unknown $\theta$.

Under this specification, we use the Gaussian density for $\alpha$, and the posterior becomes
\begin{align}
    \pi\big(y \mid \theta\,; \mathcal{H}\big) &\propto \int \exp\left(-\frac{\big(y - \left\langle \theta, z\right\rangle\big)^2}{2\sigma^2}\right) 
\prod_{j\in[J]}h_{j}(z_j)\mathrm{d}z_j\\
&= \int \exp\left(-\frac{y^2 - 2y\left\langle \theta, z\right\rangle + \left\langle \theta, z\right\rangle^2}{2\sigma^2}\right) 
\prod_{j\in[J]}h_{j}(z_j)\mathrm{d}z_j. 
\end{align}

\paragraph{Approximation under small $y$.} Lastly, we return to our original problem. In our problem, we aim to minimize the loss as possible. In the other words, our goal is to find $\theta$ such that $\left\langle \theta, z\right\rangle$ is as small as possible. 

If $y$ is sufficiently small negative value, 
\begin{align}
    \pi\big(y \mid \theta\,; \mathcal{H}\big) &\propto  \int \exp\left(-\frac{y^2 - 2y\left\langle \theta, z\right\rangle + \left\langle \theta, z\right\rangle^2}{2\sigma^2}\right) 
\prod_{j\in[J]}h_{j}(z_j)\mathrm{d}z_j\\
&\propto \int \exp\left(-\left\langle \theta, z\right\rangle\right) 
\prod_{j\in[J]}h_{j}(z_j)\mathrm{d}z_j,
\end{align}
where we ignored the term $y^2$ that is irrelevant to the optimization of $\theta$ and the term $\left\langle \theta, z\right\rangle^2$, which is relatively insignificant compared to $-y\left\langle \theta, z\right\rangle$ as $y$ is sufficiently small.

\section{Dynamic GBPS}
\label{sec:time_series}
This section provides an extension of the basic GBPS to a dynamic setting. We consider time-series data and decision-making with sequential observations. One of the most important applications in this formulation is portfolio optimization, and we focus on this application in our explanation. 

\subsection{Problem Setting}
We consider a time series with length $T$. In each period $t \in [T]$, we assume the availability of $J$ experts who provide predictive distributions of their own policy $w_{t,j} = (w_{t,j}(a))_{a \in [K]}$. In each period $t$, expert $j$ incurs a loss defined as
\begin{align}
    z_{t,j} = \sum_{a \in [K]} w_{t,j}(a) \ell_t(a),
\end{align}
where $\ell_t: [K] \to \mathbb{R}$ is an $\mathcal{F}_{t-1}$-measurable loss of action $a$, $z_t \coloneqq (z_{t,j})_{j \in [J]}$, $z_{1:t} \coloneqq (z_s)_{s \in [t]}$, and $\mathcal{F}_{t-1} \coloneqq \sigma(z_{1:t-1})$.

The predictive distribution of $z_{t,j}$ for expert $j$ is denoted by $h_{t,j}(z_{t,j})$. We also define $\mathcal{H}_{1:t} \coloneqq (h_{s,j})_{j \in [J], s \in [t]}$. 

Let $\theta_t \in \Delta^J$ be an ensemble parameter, and let $\pi^{\theta_t, w}_t$ be a new policy in period $t$ defined as
\begin{align}
    \pi^{\theta_t, w}_t(a) = \sum_{j \in [J]} \theta_{t,j} w_{t,j}(a).
\end{align}

Under this new policy $\pi^{\theta_t, w}_t(a)$, we incur the following loss:
\begin{align}
    \sum_{j \in [J]} \theta_{t,j} \sum_{a \in [K]} w_{t,j}(a) \ell_t(a) = \sum_{j \in [J]} \theta_{t,j} z_{t,j},
\end{align}
which is denoted by $L_t(\theta_t, z_t)$.

\begin{example}[Portfolio optimization]
Consider portfolio optimization in finance. Assume that there are $K$ financial assets. In each period $t$, each expert $j$ recommends an asset allocation ratio $w_{t,j} \in \Delta^K$, under which we incur a loss $\sum_{a \in [K]} w_{t,j}(a) \ell_t(a)$. We can define the loss as the negative return of the asset allocation. By utilizing the experts' advice, we conduct asset allocation with an allocation ratio $\sum_{j \in [J]} \theta_{t,j} w_{t,j}(a)$ for each asset $a \in [K]$. Our goal is to obtain $\theta_{t,j}$ to minimize the cumulative loss $\sum_{t \in [T]} \sum_{a \in [K]} w_{t,j}(a) \ell_t(a)$. 
\end{example}

\subsection{Dynamic GBPS}
In dynamic GBPS, in addition to the basic formulation of GBPS, we also consider the state-space transition, which captures the uncertainty of the time series. 

In dynamic GBPS, we consider the following posterior distribution of $\theta_{t+1}$:
\begin{align}
    \pi(\theta_{t+1} \mid \mathcal{H}_{1:t+1}) &= \int \pi(\theta_{t+1} \mid \theta_t) \pi(\theta_t \mid \mathcal{H}_{1:t+1}) \, \mathrm{d} \theta_t,
\end{align}
where
\begin{align*}
       \pi(\theta_t \mid z_{1:t}, \mathcal{H}_{1:t+1}) &\propto \int \cdots \int \exp\left(-L(\theta_t, z_{t,j})\right) \prod_{j \in [J]} h_j(z_{t,j}) \, \mathrm{d}z_{t,j},
\end{align*}
$\pi(\theta_{t+1} \mid \theta_t)$ is the conditional density of $\theta_{t+1}$ under a linear model $\theta_{t+1} = \theta_t + \omega_t$ with $\omega_t \sim \mathcal{N}(0_J, W_t)$, and $0_J$ and $W_t$ are a $J$-dimensional zero vector and a positive semi-definite ($J \times J$)-matrix, respectively. 

\subsection{Justification}
Our formulation is inspired by dynamic linear models \citep{WestHarrison1997book2} and dynamic BPS \citep{mcalinn2019dynamic}.
\paragraph{Dynamic BPS.}
We first recap the dynamic BPS proposed by \citet{mcalinn2019dynamic}. Instead of a decision-making problem, we consider the problem of predicting $y_t \in \mathbb{R}$. 

Under the dynamic BPS framework, we compute the posterior of $y_t$ as 
\begin{align}
    \pi(y_t \mid y_{1:t-1}, \mathcal{H}_{1:t}) \coloneqq \int \pi(y_t \mid \theta_t, y_{1:t-1}, \mathcal{H}_{1:t}) \pi(\theta_t \mid y_{1:t-1}, \mathcal{H}_{1:t}) \, \mathrm{d} \theta_t,
\end{align}
where $y_{1:t-1} = (y_s)_{s \in [t-1]}$, 
\begin{align}
    \pi(y_t \mid \theta_t, y_{1:t-1}, \mathcal{H}_{1:t}) \coloneqq \pi(y_t \mid \theta_t, \mathcal{H}_{1:t}) = \int \alpha_t(y_t \mid z_t, \theta_t) \prod_{j \in [J]} h_{t,j}(z_{t,j}) \, \mathrm{d} z_{t,j},\\
    \pi(\theta_t \mid y_{1:t-1}, \mathcal{H}_{1:t}) \coloneqq \int \pi(\theta_t \mid \theta_{t-1}) \pi(\theta_{t-1} \mid y_{1:t-1}, \mathcal{H}_{1:t}) \, \mathrm{d} \theta_{t-1},\\
    \pi(\theta_{t-1} \mid y_{1:t-1}, \mathcal{H}_{1:t}) \coloneqq \frac{\pi(y_{t-1} \mid \theta_{t-1}, \mathcal{H}_{1:t}) \pi(\theta_{t-1} \mid y_{1:t-2}, \mathcal{H}_{1:t})}{\pi(y_{t-1} \mid y_{1:t-2})},
\end{align}
and $\alpha_t(y_t \mid z_t, \theta_t)$ is a synthesis function that synthesizes the experts' policies (losses). 

\paragraph{Dynamic Linear Models for BPS.}
Various definitions can be given to the synthesis function $\alpha_t$, but in this study, we focus on \emph{dynamic linear models} following \citep{mcalinn2019dynamic} and \citep{mcalinn2020multivariate}, as
\begin{align}
    &y_t = z^\top_t \theta_t + \nu_t, \quad \nu_t \sim \mathcal{N}(0, \sigma^2_t),\nonumber\\
    \label{eq:time_varying_beta}
    &\theta_t = \theta_{t-1} + \omega_t, \quad \omega_t \sim \mathcal{N}(0, W_t).
\end{align}

This model is a type of state-space model and is considered suitable for modeling time series data, as addressed in this study. Then, the synthesis function can be rewritten as
\begin{align*}
    \alpha_t(y_t \mid z_t, \theta_t) = \mathcal{N}\left(y_t; z^\top_t \theta_t\right) \propto \exp\left(-\frac{y_t^2 - 2y_t \left\langle \theta_t, z_t \right\rangle + \left\langle \theta_t, z_t \right\rangle^2}{2\sigma^2_t}\right).
\end{align*}

Under this specification, we use the Gaussian density for $\alpha$, and the posterior becomes
\begin{align}
    \pi\big(y_t \mid \theta_t; \mathcal{H}_{1:t}\big) &\propto \int \exp\left(-\frac{y_t^2 - 2y_t\left\langle \theta_t, z_t\right\rangle + \left\langle \theta_t, z_t\right\rangle^2}{2\sigma_t^2}\right) 
    \prod_{j \in [J]} h_{t,j}(z_{t,j}) \, \mathrm{d} z_{t,j}.
\end{align}

\paragraph{Approximation under small $y$.} Lastly, we return to our original problem. As in the justification of the basic formulation in Section~\ref{sec:justify}, we aim to find $\theta$ such that $\left\langle \theta, z\right\rangle$ is as small as possible. If $y_t$ is sufficiently small and negative, 
\begin{align}
    \pi\big(y_t \mid \theta_t; \mathcal{H}_{1:t}\big) \propto \int \exp\left(-\left\langle \theta_t, z_t\right\rangle\right) 
    \prod_{j \in [J]} h_j(z_{t,j}) \, \mathrm{d} z_{t,j},
\end{align}
where we ignore the term $y_t^2$ that is irrelevant to the optimization of $\theta_t$ and the term $\left\langle \theta_t, z_t\right\rangle^2$, which is relatively insignificant compared to $-y_t\left\langle \theta_t, z_t\right\rangle$ as $y_t$ is sufficiently small.

Under this dynamic linear model, the time-varying coefficient $\theta_t$ follows a random walk defined by \eqref{eq:time_varying_beta}. Here, $W_t$ is defined via a standard single discount factor specification (Section 6.3 in \cite{WestHarrison1997book2}; Section 4.3 in \cite{Prado2010}), using a state evolution discount factor $e \in (0,1]$. Moreover, the residual variance $\varepsilon_t$ follows a standard beta-gamma random walk volatility model (Section 10.8 in \cite{WestHarrison1997book2}; Section 4.3 in \cite{Prado2010}), with $\varepsilon_t = \varepsilon_{t-1} \delta / \gamma_t$ for some discount factor $\delta \in (0,1]$ and where $\gamma_t$ are beta distributed innovations, independent over time and independent of $\nu_s$ and $\eta_{1,r}, \dots, \eta_{J,r}$ for all $t,s,r$. Given choices of discount factors underlying these two components, and a (conjugate normal/inverse-gamma) prior for $(w_{0,0}, w_{1,0}, \ldots, w_{J,0}, \nu_0)$ at $t=0$, the model is specified.

In BPS, since the posterior distribution cannot be obtained analytically, it is computed by simulation using Markov Chain Monte Carlo (MCMC). The details of MCMC are described in \cite{mcalinn2020multivariate}.

\section{Application to Portfolio Optimization}
\label{sec:application}
This study conducts empirical analyses using the US and Japanese stock return datasets. In each dataset, we use 10 types of stocks listed in Tables~\ref{tab:us} and \ref{tab:jap}.

We consider monthly returns. Using the data from January 1, 2009, to December 31, 2011, we estimate the portfolio weights for each method. Then, using the datasets from 2012 to 2019, we test the performance of each method. Furthermore, we sequentially update the weights during the test period.

\begin{table}[t]
    \caption{US stock data}
    \label{tab:us}
    \centering
    \scalebox{0.85}{
    \begin{tabular}{|c|c|}
    \hline
    Company & Industry \\
    \hline
Apple Inc. & Technology \\ 
Microsoft Corp. & Technology \\ 
Amazon.com Inc. & Consumer Discretionary \\ 
Alphabet Inc. & Communication Services \\ 
Berkshire Hathaway Inc. & Financials (Diversified Holdings) \\ 
Johnson \& Johnson & Health Care \\ 
Walmart Inc. & Consumer Staples (Retail) \\ 
ExxonMobil Corp. & Energy (Oil and Gas) \\ 
Procter \& Gamble Co. & Consumer Staples (Consumer Goods) \\
Intel Corp. & Technology (Semiconductors) \\ 
\hline
    \end{tabular}
}
\end{table}

\begin{table}[t]
    \caption{Japanese stock data}
    \label{tab:jap}
    \centering
    \scalebox{0.85}{
    \begin{tabular}{|c|c|}
    \hline
    Company & Industry \\
    \hline
Toyota Motor & Automotive \\
SoftBank Group & Telecommunication \& IT \\
Keyence & Electronic Equipment \\ 
Nidec Corporation & Electrical Equipment \\
Nintendo & Entertainment \\ 
Tokyo Electron & Semiconductor Manufacturing Equipment \\ 
Fast Retailing & Retail (Apparel) \\ 
Tokio Marine Holdings & Insurance \\
Astellas Pharma & Pharmaceuticals \\ 
Seven \& i Holdings & Retail (General) \\
\hline
    \end{tabular}
}
\end{table}

\paragraph{Experts}
As defined in previous sections, the experts are predictive models for asset returns $\bm{X}_t$. In this empirical studies, the asset return $\bm{X}_t$ is predicted using the following methods:
\begin{itemize}
    \item The sample mean of the past 1 year ($Mean_t[1]$).
    \item The sample mean of the past 3 years ($Mean_t[3]$).
    \item An AR$(1)$ regression model using samples from the past 3 years ($AR_t(1)$).
    \item An AR$(2)$ regression model using samples from the past 3 years ($AR_t(2)$).
    \item An AR$(3)$ regression model using samples from the past 3 years ($AR_t(3)$).
\end{itemize}

\paragraph{Alternative methods}
We compare our proposed GBPS with the experts' methods and the mean-variance (MV) portfolio with BPS. 

In the MV portfolio with BPS, we construct the MV portfolios using the BPS with the experts' predictions mentioned above. Among these MV portfolios, we choose the one with the highest Sharpe ratio.

\begin{figure*}[t]\centering
\includegraphics[width=120mm]{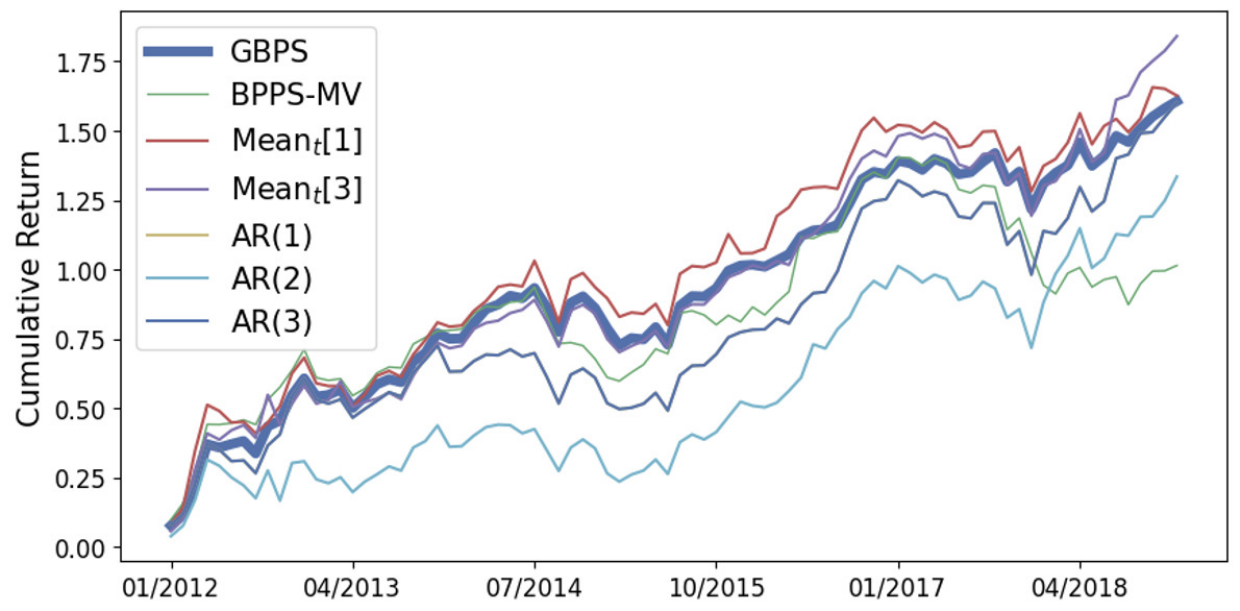}
\caption{Experimental results with US stocks. The $y$-axis in the figures represents the cumulative returns, while the $x$-axis represents the months and years.}
\label{fig:fig1}

\includegraphics[width=120mm]{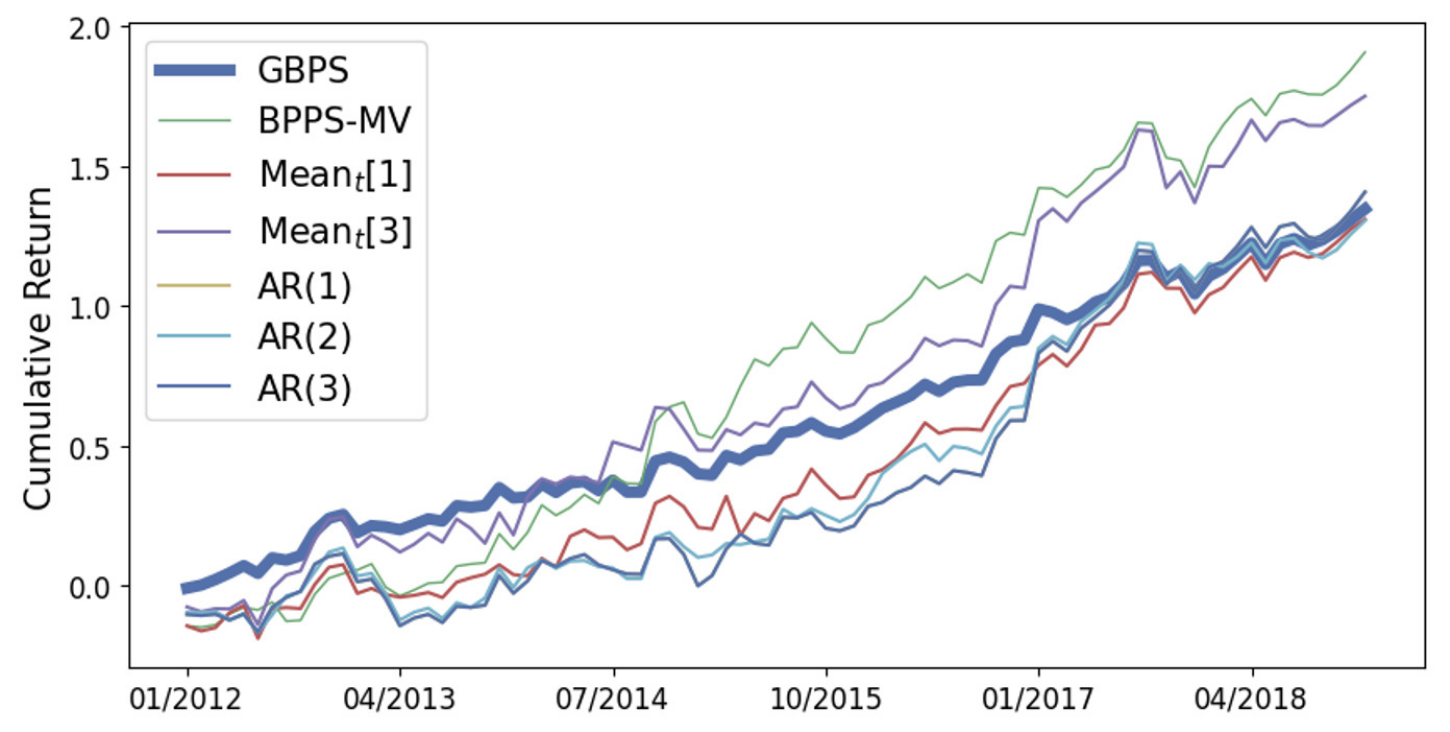}
\caption{Experimental results with Japanese stocks. The $y$-axis in the figures represents the cumulative returns, while the $x$-axis represents the months and years.}
\label{fig:fig2}
\end{figure*}

\paragraph{Experimental results}
We measure the performance of the methods by using cumulative returns. For each method, we report the cumulative returns from January 1, 2012, to December 31, 2019. We assume no cost is incurred by changing the weights. We present the results for the US market in Figure~\ref{fig:fig1} and for the Japanese market in Figure~\ref{fig:fig2}.

In the empirical studies, the GBPS shows relatively high performance compared with the other methods. Furthermore, its performance does not drop significantly across different periods. From these results, we conclude that the GBPS is a high-performance and stable method, although it is not the best among the candidates.

\section{Conclusion}
In this study, we proposed the GBPS, a novel Bayesian framework for decision-making that integrates experts' predictive distributions to improve decision quality by minimizing expected loss. Unlike existing methods that rely on parameter estimation, GBPS employs a loss minimization strategy inspired by general Bayes principles. We extended this framework to dynamic settings, accommodating sequential decision-making with time-series data through state-space models, making it suitable for applications like portfolio optimization. Simulation studies demonstrated the effectiveness of GBPS in real-world scenarios.

\bibliographystyle{icml2024}
\bibliography{arXiv.bbl}

\end{document}